%% file: paper.tex
\documentclass[3p,review]{elsarticle}

\usepackage{acronym}
\usepackage{hyperref}
\usepackage{subfig}
\usepackage{amsmath}
\usepackage{booktabs}
\usepackage{graphicx}
\usepackage{marginnote}
\usepackage{marginnote}
\usepackage{hyperref}
\usepackage{subfig}
\usepackage{caption}
\usepackage{hyperref}
\usepackage{xcolor}

\graphicspath{{.}{images/}}
\DeclareGraphicsExtensions{.eps}

\makeatletter
\AtBeginDocument{%
  \renewcommand*{\AC@hyperlink}[2]{#2}%
}
\makeatother

\journal{Simulation Modelling Practice and Theory}

\begin{document}

\begin{frontmatter}
  \title{Fault Tolerant Adaptive Parallel and Distributed Simulation through Functional Replication \footnotemark[0] \footnotemark[1]}

\footnotetext[0]{An early version of this work appeared in~\cite{gda-dsrt-2016}. This paper is an extensively revised and extended version of the previous work in which more than 30\% is new material.}

\footnotetext[1]{The publisher version of this paper is available at \url{https://doi.org/10.1016/j.simpat.2018.09.012}.
\textbf{{\color{red}Please cite this paper as: ``Fault Tolerant Adaptive Parallel and Distributed Simulation through Functional Replication. Simulation Modelling Practice and Theory, vol. 93 (May 2019), Elsevier''.}}}

\author{Gabriele D'Angelo\corref{cor1}}
\cortext[cor1]{Corresponding Author. Address: Department of Computer Science and Engineering. University of Bologna. Mura Anteo Zamboni 7. I-40127, Bologna. Italy. Phone +39 0547 338886, Fax +39 051 2094510} 
\ead{g.dangelo@unibo.it}
\author{Stefano Ferretti}
\ead{s.ferretti@unibo.it}
\author{Moreno Marzolla}
\ead{moreno.marzolla@unibo.it}
\address{Department of Computer Science and Engineering\\University of Bologna, Italy}

\begin{abstract}
This paper presents FT-GAIA, a software-based fault-tolerant parallel and distributed simulation middleware.
FT-GAIA has being designed to reliably handle Parallel And Distributed Simulation (PADS) models, 
which are needed to properly simulate and analyze complex systems arising in any kind of scientific 
or engineering field.
PADS takes advantage of multiple execution units run in multicore processors, cluster of workstations or HPC systems. 
However, large computing systems, such as HPC systems that include hundreds of thousands of 
computing nodes, have to handle frequent failures of some components. 
To cope with this issue, FT-GAIA transparently replicates simulation entities and
distributes them on multiple execution nodes. This allows the simulation to tolerate crash-failures of computing nodes.
Moreover, FT-GAIA offers some protection against Byzantine failures, since interaction messages among the simulated entities are replicated as well, 
so that the receiving entity can identify and discard corrupted messages.
Results from an analytical model and from an experimental evaluation show that FT-GAIA provides a high degree of
fault tolerance, at the cost of a moderate increase in the computational load of the execution units.
\end{abstract}


\end{frontmatter}

\input{sec_introduction.tex}
\input{sec_related-work.tex}
\input{sec_gaia-artis.tex}
\input{sec_fault-tolerant-simulation.tex} 
\input{sec_experimental-evaluation.tex}
\input{sec_analytical-evaluation.tex}
\input{sec_conclusions.tex}

\section*{Symbols}

\begin{tabular}{rp{.75\textwidth}}
  $L :=$ & Number of Logical Processes (LPs)\\
  $N :=$ & Number of Simulation Entities (SEs)\\
  $M :=$ & Number of copies of each SE ($M \in \{0, \ldots, L\}$)\\
  $X :=$ & Number of crashed LPs ($X \in \{0, \ldots, L\}$)\\
  $N_i :=$ & Number of instances of SEs $i$ that do not crash\\
  $R_C :=$ & System reliability under the crash failure model\\
  $R^*_C :=$ & System reliability under the crash failure model (without the FT-GAIA constraint)\\
  $R_B :=$ & System reliability under the Byzantine failure model
\end{tabular}

\section*{Acronyms}

\begin{acronym}[PDES]
  \acro{DES}{Discrete Event Simulation}
  \acro{FEL}{Future Event List}
  \acro{GVT}{Global Virtual Time}
  \acro{HPC}{High Performance Computing}
  \acro{IRP}{Inertial Reference Platform}
  \acro{LVT}{Local Virtual Time}
  \acro{LP}{Logical Process}
  \acroplural{LP}[LPs]{Logical Processes}
  \acro{MTTF}{Mean Time To Failure}
  \acro{PADS}{Parallel And Distributed Simulation}
  \acro{PDES}{Parallel Discrete Event Simulation} 
  \acro{PE}{Processing Element}
  \acro{SE}{Simulated Entity}
  \acroplural{SE}[SEs]{Simulated Entities}
  \acro{WCT}{Wall Clock Time}
\end{acronym}


\bibliographystyle{elsarticle-num}
\bibliography{paper}

\end{document}

%% file: sec_introduction.tex
\section{Introduction}\label{sec:introduction}

Computer simulation is an important tool to model, analyze and
understand physical, biological and social phenomena. Among the
different methodologies \ac{DES} is of particular interest, since it
is frequently employed to model and analyze many types of systems,
including computer architectures, communication networks, street
traffic and others.

In a~\ac{DES}, the system is modeled as a set of entities that
interact.  The simulation has a state which evolves through the
generation of \emph{events} issued by simulated entities or by a
(human or synthetic) supervisor of the simulation.  Events occur at
discrete points in time. The overall structure of a sequential
event-based simulator is relatively simple: the simulator engine
maintains a list, called~\ac{FEL}, of all pending events, sorted in
non decreasing time of occurrence. The execution of the simulation
consists of a loop: at each iteration, the event with lower
timestamp~$t$ is removed from the~\ac{FEL}, and the simulation time is
advanced to~$t$. Then, the event is executed, possibly triggering the
generation of new events to be scheduled for execution at some future
time.

Continuous advances in our understanding of complex systems, combined
with the need for higher model accuracy, demand an increasing amount
of computational power.  The simulation of complex systems might
generate a huge amount of events, due to the enormous amount of
entities to be simulated and the high rate of events they trigger.
Just as an example, think at the Internet of Things (IoT), the network
of physical devices, vehicles, home appliances and other items
embedded with computational and that communication capabilities, that
nowadays is considered the most prominent infrastructure on top of
which novel smart services will be implemented.  Simulating such a
kind of system is very demanding and imposes the use of sophisticated
simulation techniques~\cite{gda-simpat-iot}.  In this kind of
scenarios, sequential~\ac{DES} techniques become inappropriate for
analyzing large or detailed models.  \ac{DES} must thus evolve into
something that is able to handle simulations at larger scales.

An alternative approach, called~\ac{PDES} refers to the execution of a
single discrete event simulation program on a parallel
computer~\cite{Fujimoto:1990:PDE:84537.84545}. The goal is to
parallelize the execution of the simulation events for better
scalability.

\ac{PADS} is concerned with the execution of a simulation program on
computing platforms containing multiple processors~\cite{Fuj00}.
\ac{PADS} takes advantage of multiple execution units to efficiently
handle large simulation models. These execution units can be
distributed across the Internet, or grouped as massively parallel
computers or multicore processors.  While~\ac{PADS} has been used for
concurrent execution of many different simulation paradigms 
(e.g.~continuous simulation, concurrent replication), this paper focuses on
the distributed execution of discrete event simulations, i.e.~we use
the~\ac{PADS} techniques for implementing~\ac{DES} models.

\begin{figure}[t]
  \centering\includegraphics[scale=.7]{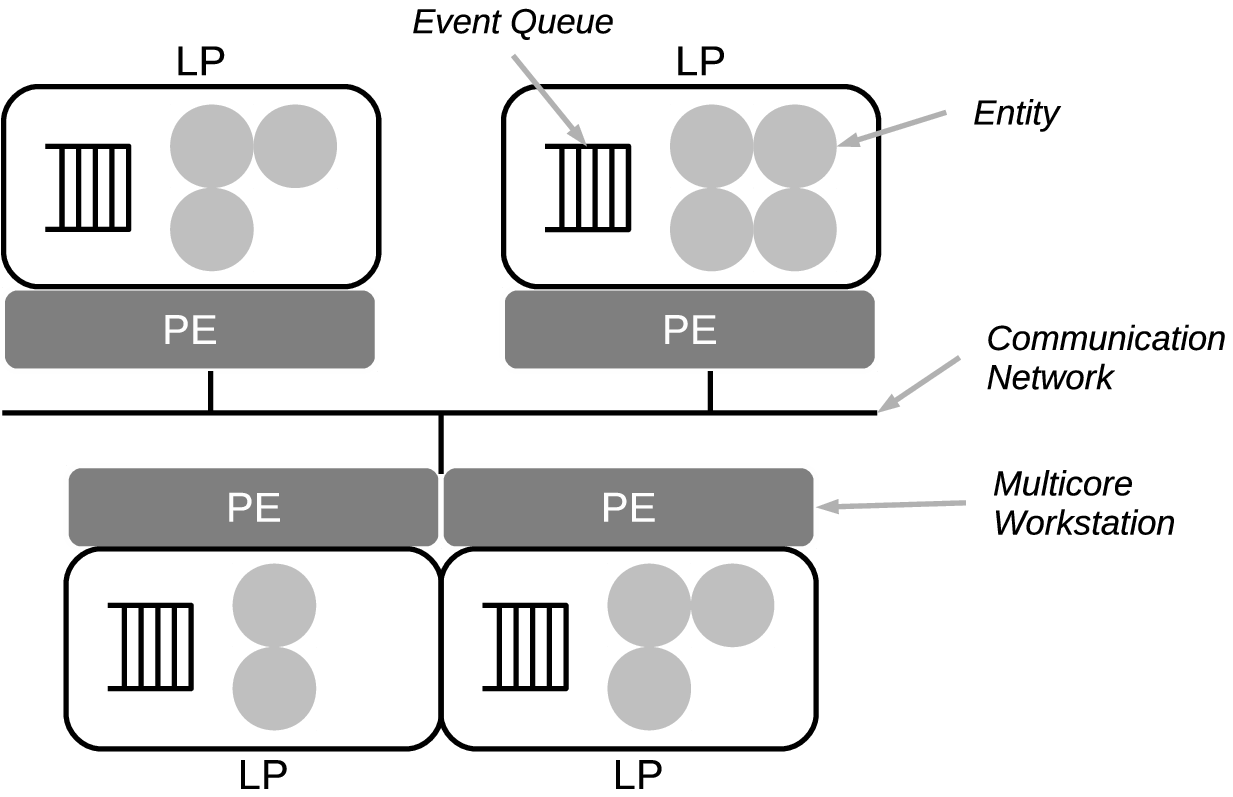}
  \caption{
  Structure of a~\acl{PADS} that implements a~\acl{DES} model.}\label{fig:pads}
\end{figure}

More in detail, in~\ac{PADS}, the simulation model is partitioned in
submodels, called~\acp{LP} which can be evaluated concurrently by
different~\acp{PE}. More precisely, the simulation model is described
in terms of multiple interacting~\acp{SE} which are assigned to
different~\acp{LP}. Each~\ac{LP} runs on a different~\ac{PE}, where
a~\ac{PE} is an execution unit acting as a container of a set of
entities.  The simulation execution consists of the exchange of
timestamped messages, representing simulation events, between
entities. Each~\ac{LP} has an incoming queue where messages are
inserted before being dispatched to the appropriate
entities. Without loss of generality, through this paper we
  will assume that a~\ac{PE} is a single core of a multicore
  processor. Figure~\ref{fig:pads} shows the general structure of a
parallel and distributed simulator.

Clearly enough, \ac{PADS} can strongly benefit from the use of cloud
computing infrastructures. Cloud computing allows instantiating and
dynamically maintaining computing (virtual) machines that meet
arbitrarily varying resource requirements. Service level agreements
can be employed in order to understand if the cloud provides the
Quality-of-Service the user is expecting~\cite{Ferretti:2010}.  QoS
guarantees, together with the possibility of arbitrarily adding or
removing resources on demand, provide the simulationist with a very
useful computing environment to execute complex simulations, without
having to manage the computing infrastructure~\cite{Marzolla:2012}.
However, as in every distributed system, cloud virtual machines can
fail. Thus, fault tolerance schemes are
required~\cite{Fujimoto:2016:RCP:2892241.2866577}.

Execution of long-running applications on increasingly larger parallel
machines is likely to hit the \emph{reliability
wall}~\cite{reliability-wall}. This means that, as the system size
(number of components) increases, so does the probability that at
least one of those components fails, therefore reducing the
system~\ac{MTTF}. At some point the execution time of the parallel
application may become larger than the~\ac{MTTF} of its execution
environment, so that the application has little chance to terminate
normally.

As a purely illustrative example, let us consider a \ac{PADS} with
$L$~\acp{LP}. Let~$X_i$ be the stochastic variable representing the
duration of uninterrupted operation of the $i$-th~\ac{LP}, $1 \leq i
\leq L$, taking into account both hardware and software failures. For
the sake of simplicity, we assume that each~\ac{LP} resides on a
different~\ac{PE}, so that each hardware failure (i.e.~a~\ac{PE}
crash) affects an~\ac{LP} only. Assuming that all~$X_i$ are
independent and exponentially distributed (this assumption is somewhat
unrealistic but widely used~\cite{bolch}), we have that the
probability $P(X_i > t)$ that~\ac{LP} $i$ operates without failures
for at least~$t$ time units is
\[
P(X_i > t) = e^{-\lambda t}
\]
\noindent where $\lambda$ is the failure rate. The joint probability
that all $L$~\acp{LP} operate without failures for at least~$t$ time
units is therefore $R(L, t) = \prod_i P(X_i > t) = e^{-L \lambda t}$;
this is the formula for the reliability of~$L$ components connected in
series, where each component fails independently, and a single failure
brings down the whole system.

\begin{figure}[t]
  \centering%
  \includegraphics[width=.8\textwidth]{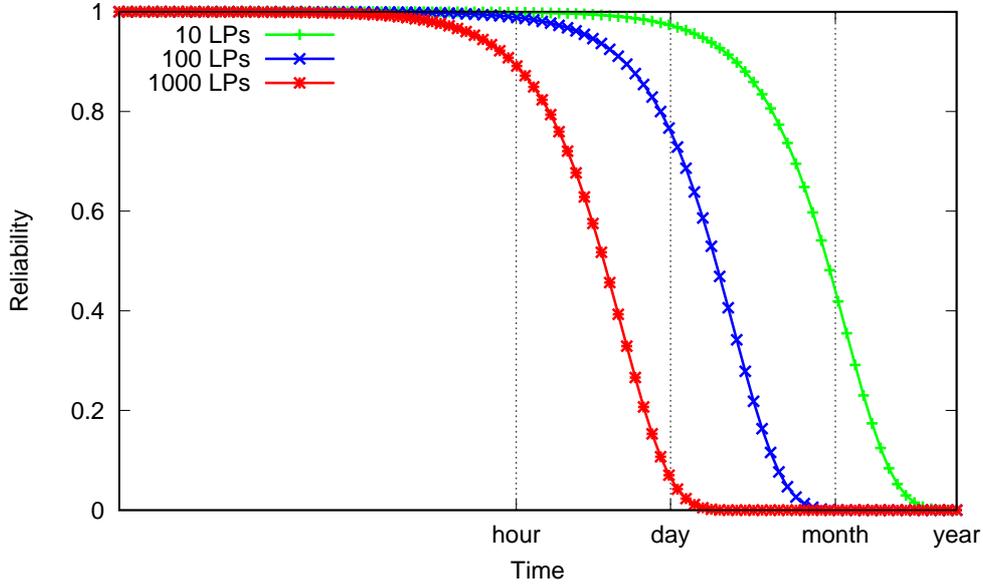}
  \caption{System reliability of parallel and distributed simulation
    with different number of~\acp{LP}, assuming that the~MTTF for
    each~\ac{LP} is one year; higher is better, log scale on the
    horizontal axis.}\label{fig:reliability}
\end{figure}

Figure~\ref{fig:reliability} shows the value of~$R(L, t)$ (the
probability of no failures for at least~$t$ consecutive time units)
for systems with $L=10, 100, 1000$ \acp{LP}, assuming a~\ac{MTTF} of
one year ($\lambda \approx 2.7573 \times 10^{-8} s^{-1}$). We can see
that the system reliability quickly drops as the number of~\acp{LP}
increases: a simulation involving $L=1000$ \acp{LP} and requiring one
day to complete is very unlikely to terminate successfully. 

Although the model above is overly simplified, and is not intended to
provide an accurate estimate of the reliability of actual \ac{PADS},
it does show that building a reliable system out of a
large number of unreliable parts is challenging.  

\begin{table}[t]
  \centering\scriptsize%
  \begin{tabular}{llrrr}
    \toprule
    Name      & System                  & N. of cores  & $R_\textrm{max}$ & $R_\textrm{peak}$ \\
              &                         &              & (TFlop/s)   & (TFlop/s)\\
    \midrule
    Summit    & IBM Power System AC922  &  $2,282,544$ & $122,300.0$ & $187,659.3$\\
    Sunway TaihuLight & Sunway MPP      & $10,649,600$ &  $93,014.6$ & $125,435.9$\\
    Sierra    & IBM Power System S922LC &  $1,572,480$ &  $71,610.0$ & $119,193.6$\\
    Tianhe-2A & TH-IVB-FEP Cluster      &  $4,981,760$ &  $61,444.5$ & $100,678.7$\\ 
    ABCI      & PRIMERGY CX2550 M4      &    $391,680$ &  $19,880.0$ &  $32,576.6$\\
    Piz Daint & Cray XC50               &    $361,760$ &  $19,590.0$ &  $25,326.3$\\
    Titan     & Cray XK7                &    $560,640$ &  $17,590.0$ &  $27,112.5$\\
    Sequoia   & BlueGene/Q              &  $1,572,864$ &  $17,173.2$ &  $20,132.7$\\
    Trinity   & Cray XC40               &    $979,968$ &  $14,137.3$ &  $43,902.6$\\
    Cori      & Cray XC40               &    $622,336$ &  $14,014.7$ &  $27,880.7$\\
    \bottomrule
  \end{tabular}
  \caption{The top ten HPC systems in June 2018 Top500 Supercomputer
    list. $R_\textrm{max}$ and $R_\textrm{peak}$ are the maximum and
    theoretical peak LAPACK performance,
    respectively.}\label{tab:top500}
\end{table}

  To put the numbers above more in context, we report on
  Table~\ref{tab:top500} the number of cores in the top ten~\ac{HPC}
  systems that appear on the June 2018 edition of the Top500
  Supercomputer list\footnote{\url{https://top500.org/lists/2018/06/},
    accessed August, 2018}. Five systems (Summit, Sunway TaihuLight,
  Sierra, Tianhe-2A, and Sequoia) have more than one million cores,
  while the others are in the range of hundreds of thousands.  As the
  size of~\ac{HPC} systems grows, reliabilty issues become more and
  more relevant~\cite{Yang:2012}.

  The reliability of~\ac{HPC} systems has been investigated, among
  others, in~\cite{Schroeder:2010, ElSayed:2013}.
  In~\cite{Schroeder:2010}, the authors report about~$0.5$ hardware
  failures/year per processor on average, across several
  different~\ac{HPC} systems. It is quite instructive to observe that
  the root cause of faults include environmental factors (e.g., power
  outages), human errors, network failures, software errors, and
  hardware failures~\cite{ElSayed:2013, Egwutuoha:2013}.

  Therefore, a~$10$-million processors \ac{HPC} system with
  a~\ac{MTTF} of two years for each processor will experience $10^7 /
  2 = 5 \times 10^6$ failures/year. In general, it is well understood
  that no matter how reliable the basic components are, the future
  generation of supercomputers will experience an ever increasing
  stream of failures and must cope with them~\cite{reliability-wall}.

This paper describes a novel approach to deal with fault tolerance in \ac{PADS}.
The proposed solution, termed FT-GAIA, is a fault tolerant extension of the
GAIA/ART\`IS parallel and distributed simulation middleware~\cite{gda-perf-2005,gda-simpat-2017}. 
FT-GAIA deals with crash errors 
and Byzantine faults by resorting to \emph{server groups}~\cite{cristian93}: simulation
entities are replicated, in the cloud / distributed computing system, so that the model can be executed even if some
of them fail. This functional replication is implemented by adding 
a related software layer in the GAIA/ART\`IS stack. The replication of all the simulated entities
is transparent to user-level.
Thus, FT-GAIA can be used as a drop-in replacement to GAIA/ART\`IS
when fault tolerance is the major concern. Needless to say, 
fault tolerance increases the computational and communication loads at \acp{LP}, 
thus causing a moderate increment on the performance of the simulator.

The remainder paper is organized as follows. In Section~\ref{sec:related-work}
we review the state of the art related to fault tolerance in~\ac{PADS}. The
GAIA/ART\`IS parallel and distributed simulation middleware is
described in Section~\ref{sec:gaia-artis}.
Section~\ref{sec:fault-tolerant-simulation} is devoted to the
description of FT-GAIA, a fault tolerant extension to GAIA/ART\`IS.
An empirical performance evaluation of FT-GAIA, based on a prototype
implementation that we have developed, is discussed in
Section~\ref{sec:experimental-evaluation}. 
Section \ref{sec:analytical-evaluation} discusses a probabilistic model that drives an analytical evaluation of the proposed scheme.
Finally, Section~\ref{sec:conclusions} provides some concluding remarks.

%% file: sec_related-work.tex
\section{Background and Related Work}\label{sec:related-work}

In distributed systems, two typical approaches used to cope with 
hardware-related reliability are \emph{checkpointing} and 
\emph{functional replication}. 

The checkpoint-restore paradigm
requires the running application to periodically save its state on
non-volatile storage (e.g.~disk) so that it can resume execution from
the last saved snapshot in case of failure. It should be observed that
saving a snapshot may require considerable time; therefore, the
interval between checkpoints must be carefully tuned to minimize the
overhead.

Functional replication consists of replicating parts of the
application on different execution nodes, so that failures can be
tolerated if there is some minimum number of running instances of each
component. Note that each component must be modified so that it is
made aware that multiple copies of its peers exist, and can interact
with all instances appropriately.

It is important to remark that functional replication is not effective
against logical errors, i.e., bugs in the running applications, since
the bug can be triggered at the same time on all instances. A
prominent -- and frequently mentioned -- example is the failure of the
Ariane 5 rocket that was caused by a software error on
its~\acp{IRP}. There were two~\ac{IRP}, providing hardware
fault tolerance, but both used the same software. When the two
software instances were fed with the same (correct) input from the
hardware, the bug (an uncaught data conversion exception) caused both
programs to crash, leaving the rocket without
guidance~\cite{ariane-5}. The $N$-version programming
technique~\cite{n-version} can be used to protect against software
errors, and requires running several functionally equivalent programs
that have been independently developed from the same specifications.

Although fault tolerance is an important and widely discussed topic in
the context of distributed systems research, it received comparatively
little attention by the~\ac{PADS} community. 
In what follows, we describe related works on simulation that deal 
with this main issue.

\subsection{Checkpointing}

In~\cite{Damani:1998:FDS:278008.278014} the authors propose a rollback
based optimistic recovery scheme in which checkpoints are periodically
saved on stable storage. The distributed simulation uses an optimistic
synchronization scheme in which out-of-order (i.e.~``straggler'') events
are handled according to the Time Warp protocol~\cite{Jefferson85}. 
The novel idea of this approach is to model failures as
straggler events with a timestamp equal to the last saved
checkpoint. In this way, the authors can leverage the Time Warp
protocol to handle failures.

In~\cite{Eklof:2005:FFH:1162708.1162915,Eklof:2006:EFM:1136644.1136877}
the authors propose a framework called Distributed Resource
Management System (DRMS) to implement reliable IEEE 1516
federation~\cite{HLA}. The DRMS handles crash failures using
checkpoints saved to stable storage, that is then used to migrate
federates from a faulty host to a new host when necessary. The
simulation engine is again based on an optimistic synchronization
scheme, and the migration of~\acp{LP} (the so called ``federates'' in the IEEE 1516 terminology) is implemented through Web
services.

In~\cite{Chen20081487} the authors propose a decoupled federate
architecture in which each IEEE 1516 federate is separated into a
virtual federate process and a physical federate process. The former
executes the simulation model and the latter provides middleware
services at the back-end. This solution enables the implementation of
fault tolerant distributed simulation schemes through migration of
virtual federates.

The~CUMULVS middleware~\cite{Kohl:1998:EFF:281035.281042} introduces
the support for fault tolerance and migration of simulations based on
checkpointing. The middleware is not designed to support~\ac{PADS} but
it allows the migration of running tasks for load balancing and to
improve a task's locality with a required resource.

A slightly different approach is proposed in~\cite{Luthi2004}. In 
which, the authors introduce the Fault Tolerant Resource Sharing System 
(FT-RSS) framework. The goal of FT-RSS is to build fault tolerant 
IEEE 1516 federations using an architecture in which a separate FTP
server is used as a persistent storage system. The persistent storage
is used to implement the migration of federates from one node to another.
The FT-RSS middleware supports replication of federates, partial
failures and fail-stop failures.

Recently, in~\cite{LI201690} the authors proposed a transparent middleware
for dealing with Byzantine fault in HLA-based parallel and distributed simulations.
In this case, the solution is based on the usage of replication, checkpointing and message
logging technologies. 

Finally, an approach based on the usage of virtualization techniques is described in~\cite{Malik:2017:CMI:3140065.3140066}.
The authors introduce a fault resilient framework that dynamically handles virtual machines failures inside the
cloud environment. The proposed fault resilient framework is based on state saving and snapshots of processed event list
that are implemented in each~\ac{LP}.

\subsection{Functional Replication}

In~\cite{Agrawal:1992:ROT:167293.167662} the authors propose the use
of functional replication in Time Warp simulations with the aim to
increase the simulator performance and to add fault
tolerance. Specifically, the idea is to have copies of the most
frequently used simulation entities at multiple sites with the aim of
reducing message traffic and communication delay. This approach is
used to build an optimistic fault tolerance scheme in which it is
assumed that the objects are fault free most of the time. The rollback
capabilities of Time Warp are then used to correct intermittent and
permanent faults.

In~\cite{Liris-4840} the authors describe DARX, an adaptive
replication mechanism for building reliable multi-agent systems. Being
targeted to multi-agent systems, rather than~\ac{PADS}, DARX is mostly
concerned with adaptability: agents may change their behavior at any
time, and new agents may join or leave the system. Therefore, DARX
tries to dynamically identify which agents are more ``important'', and
what degree of replication should be used for those agents in order to
achieve the desired level of fault tolerance. It should be observed
that DARX only handles crash failures, while FT-GAIA also deals with
Byzantine faults.

%% file: sec_gaia-artis.tex
\section{The GAIA/ART\`IS Middleware}\label{sec:gaia-artis}

To make this paper self-contained, we provide in this section a brief
introduction of the GAIA/ART\`IS parallel and distributed simulation
middleware; the interested reader is referred to~\cite{gda-perf-2005}
and the software homepage~\cite{pads}.

The \textit{Advanced RTI System} (ART\`IS) is a parallel and
distributed simulation middleware loosely inspired by the Runtime
Infrastructure described in the IEEE~1516 standard ``High Level
Architecture'' (HLA)~\cite{ieee1516}.  ART\`IS implements a
parallel/distributed architectures where the simulation model is
partitioned in a set of~\acp{LP}~\cite{Fuj00}. As described in
Section~\ref{sec:introduction}, the execution architecture in charge
of running the simulation is composed of interconnected~\acp{PE} and
each~\ac{PE} runs one or more~\acp{LP} (usually, a~\ac{PE} hosts
one~\ac{LP}).

In a~\ac{PADS}, the interactions between the model components are
driven by message exchanges. The low computation/communication ratio
makes~\ac{PADS} communication-bound, so that the wall-clock execution
time of distributed simulations is highly dependent on the performance
of the communication network (i.e.~latency, bandwidth and
jitter). Reducing the communication overhead can be crucial to speed
up the event processing rate of~\ac{PADS}. This can be achieved by
clustering interacting entities on the same physical host, so that
communications can happen through shared memory. 

Among the various services provided by ART\`IS, time management (i.e.,
synchronization) is fundamental for obtaining correct simulation runs
that respect the causality dependencies of events. ART\`IS supports
both conservative (Chandy-Misra-Bryant~\cite{cmb}) and optimistic
(Time Warp~\cite{Jefferson85}) synchronization algorithms. Moreover, a
distributed implementation of the time-stepped synchronization is included.

The \textit{Generic Adaptive Interaction Architecture} (GAIA)~\cite{gda-simpat-2017,pads,gda-simpat-2014}
is a software layer built on top of ART\`IS. In~GAIA,
each~\ac{LP} acts as the container of some~\acp{SE}: the simulation
model is partitioned in its basic components (the~\acp{SE}) that are
allocated among the~\acp{LP}. The system behavior is modeled by the
interactions among the~\acp{SE}; such interactions take the form of
timestamped messages that are exchanged among the entities. From the
user's point of view, a simulation model based on ART\`IS follows
a Multi Agent System (MAS) approach. In fact, each~\ac{SE} is an
autonomous agent that performs some actions (individual behavior) and
interacts with other agents in the simulation.

In most cases, the interaction between the~\acp{SE} of a~\ac{PADS} are
not completely uniform, meaning that there are clusters of~\acp{SE}
where internal interactions are more frequent. The structure of these
clusters of highly interacting entities may change over time, as the
simulation model evolves. The identification of such clusters is
important to improve the performance of a~\ac{PADS}: indeed, by
putting heavily-interacting entities on as few~\acp{LP} as possible,
we may replace most of the expensive LAN/WAN communications by more
efficient shared memory messages.

In GAIA, the analysis of the communication pattern is based on a set of 
simple self-clustering heuristics~\cite{gda-simpat-2017} that are 
provided by the framework. All the provided heuristics are generic and not model
dependent. For example, in
the default heuristic, every few timesteps for each~\ac{SE} is found
which~\ac{LP} is the destination of the large percentage
of interactions. If it is not the~\ac{LP} in which the~\ac{SE} is
contained then a migration is triggered. The migration of~\acp{SE}
among~\acp{LP} is transparent to the simulation model developer; 
entities migration is useful not only to reduce the communication
overhead, but also to achieve better load-balancing among
the~\acp{LP}, especially on heterogeneous execution platforms where
execution units are not identical. In these cases, GAIA can migrate
entities away from less powerful~\acp{PE}, towards more capable
processors if available.

%% file: sec_fault-tolerant-simulation.tex
\section{Fault Tolerant Simulation}\label{sec:fault-tolerant-simulation}

FT-GAIA is a fault tolerant extension to the GAIA/ART\`IS distributed
simulation middleware. As will be explained below, FT-GAIA uses
functional replication of simulation entities to achieve tolerance
against crashes and Byzantine failures of the~\acp{PE}.

FT-GAIA is implemented as a software layer on top of~GAIA and provides
the same functionalities of GAIA with only minor additions. Therefore,
FT-GAIA is mostly transparent to the user, meaning that any simulation
model built for~GAIA can be easily ported to FT-GAIA. The
  FT-GAIA extension will be integrated in the next release of the
  GAIA/ART\`IS simulation middleware and will be available from the
  official GAIA/ART\`IS Web site~\cite{pads}.

\begin{figure}[t]
  \centering%
  \includegraphics[scale=.8]{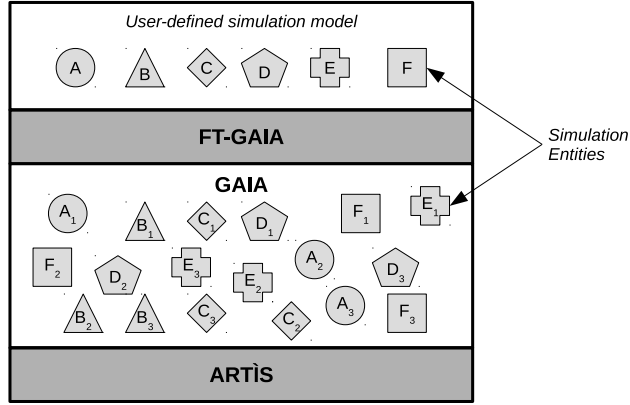}
  \caption{Layered structure of the FT-GAIA simulation engine. The
    user-defined simulation model defines a set of entities $\{A, B,
    C, D, E, F\}$; FT-GAIA creates multiple (in this example, 3)
    instances of each entity, that are handled by
    GAIA.}\label{fig:gaiaft}
\end{figure}

FT-GAIA works by replicating simulation entities (see
Fig.~\ref{fig:gaiaft}) to tolerate crash-failures and Byzantine faults
of the~\acp{LP}. A crash may be caused by a failure of the hardware --
including the network connection -- and operating system. A Byzantine
failure refers to an arbitrary behavior of a~\ac{LP} that causes
the~\ac{LP} to crash, terminate abnormally, or to send arbitrary
messages (including no messages at all) to other~\acp{LP}.

Replication is based on the following principle. If a conventional,
non-fault tolerant distributed simulation is composed of~$N$ distinct
simulation entities, FT-GAIA generates $N \times M$ entities, by
generating~$M$ independent instances of each simulation entity. All
instances $A_1, \ldots A_M$ of the same entity~$A$ perform the same
computation: if no fault occurs, they produce the same result. 

Replication comes with a cost, both in term of additional processing
power that is needed to execute all instances, and also in term of an
increased communication load between the~\acp{LP}. Indeed, if two
entities~$A$ and~$B$ communicate by sending a message from~$A$ to~$B$,
then after replication each instance $A_i$ must send the same message
to all instances $B_j$, $1 \leq i, j \leq M$, resulting in $M^2$
(redundant) messages. Therefore, the level of replication~$M$ must be
chosen wisely in order to achieve a good balance between overhead and
fault tolerance, also depending on the types of failures (crash
failures or Byzantine faults) that the user wants to address.

\paragraph*{Handling crash failures} A crash failure happens
when a~\ac{LP} crashes, but operates correctly until it halts.  When
a~\ac{LP} terminates, all simulation entities running on that~\ac{LP}
stop their execution and the local state of the computation is
lost. From the theory of distributed systems, it is known that~$M$
instances of each simulation entity are required to tolerate up
to~$(M-1)$ crash failures. Each instance must be executed on a
different~\ac{LP}, so that the failure of a~\ac{LP} only affects one
instance of all entities executed there. This is equivalent to
running~$M$ copies of a monolithic (sequential) simulation, with the
difference that a sequential simulation does not incur in
communication and synchronization overhead. However, unlike sequential
simulations, FT-GAIA can take advantage of more than~$M$ \acp{LP}, by
distributing all the $N \times M$ entities on the available execution
units. This reduces the workload on the~\acp{LP}, reducing the
wall-clock execution time of the simulation model.

\paragraph*{Handling Byzantine Failures}
Byzantine failures include all types of abnormal behaviors of
a~\ac{PE}. Examples are: the crash of a component of the distributed
simulator (e.g., \ac{LP} or entity); the transmission of
erroneous/corrupted data from an entity to other entities; computation
errors that lead to erroneous results. In this case~$M$ instances of
each~\ac{SE} are necessary to tolerate up to $\lfloor (M-1)/2 \rfloor$
Byzantine faults using the \emph{majority rule}: a~\ac{SE}
instance~$B_i$ can process an incoming message~$m$ from~$A_j$ when it
receives one copy of~$m$ from the (strict) majority of the instances
of sender~$A$ (the strict majority of~$M$ instances is $\lceil (M+1)/2
\rceil$). This applies to synchronous systems where the message delay
is bounded and faulty nodes cannot forge messages (i.e., messages are
in some sense authenticated). Again, all~$M$ instances of each~\ac{SE}
must be located on different~\acp{LP}.

In is worth noting that, GAIA (and therefore FT-GAIA) is based on a
time-stepped approach, leading to a synchronous system. Moreover, the
presence of a specific \emph{end-of-step} synchronization message that
needs to be received by all LPs represents a bound on the possible
latency for correct messages. Thus, we can conclude that FT-GAIA works
in a synchronous scenario.

  The majority rule, as implemented in FT-GAIA, requires that the
  sequences of messages produced by each working instance of the same
  simulation entity are equal, i.e.~the payload of the $i$-th message
  of each sequence is exactly the same. This comes from the fact that
  many simulation models require \emph{reproducibility} of the
  results, irrespective from the implementation details such as the
  number of~\acp{LP} used, or how entities are mapped to the~\acp{LP}.
  In turn, reproducibility requires that once started, the behavior of
  the simulation as a whole is fully deterministic. However, there
  might be scenarios where strict determinism is not required, 
  e.g.~in mixed simulations relying on Monte Carlo
  methods~\cite{Rubinstein:2016} when different execution paths are
  actually required. For such scenarios, Byzantine failures are
  difficult if not impossible to identify, because the messages
  produced by the instances of the same~\ac{SE} could be different yet
  correct. In these situations, deciding whether a message is correct
  or not would require some model-specific knowledge, if such
  knowledge exists at all. Extending FT-GAIA to allow the modeler to
  specify such knowledge is relatively straightforward, but so far we
  have not encountered any use case demanding it. 

\paragraph*{Allocation of Simulation Entities}
Once the level of replication~$M$ has been set, it is necessary to
decide where to create the~$M$ instances of each~\ac{SE}, so that the
constraint that each instance is located on a different~\ac{LP} is
met. In FT-GAIA the deployment of instances is performed during the
setup of the simulation model. In the current implementation, there is
a centralized service that keeps track of the initial location of
all~\ac{SE} instances.  When a new~\ac{SE} is created, the service
creates the appropriate number of instances according to the
redundancy model to be employed, and assigns them to the~\acp{LP} so
that all instances are located on different~\acp{LP}. Note that all
instances of the same~\ac{SE} receive the same initial seed for their
internal pseudo-random number generators; this guarantees that their
execution traces are the same, regardless of the~\ac{LP} where
execution occurs and the degree of replication. 
At the cost of some extra coordination among the~\acp{LP} even the 
initial~\acp{SE} deployment could be decentralized. This not 
challenging under the design viewpoint but would require a more complex
implementation and thus it has been left as future work.

\paragraph*{Message Handling}
We have already stated that fault tolerance through functional
replication has a cost in term of increased message load
among~\acp{SE}. Indeed, for a replication level~$M$ (i.e.,~there
are~$M$ instances of each~\ac{SE}) the number of messages exchanged
between entities grows by a factor of~$M^2$.

A consequence of message redundancy is that message filtering must be
performed to avoid that multiple copies of the same message are
processed more than once by the same~\ac{SE} instance. FT-GAIA takes
care of automatically filtering the excess messages according to the
fault model adopted; filtering is done outside of the~\ac{SE}, which
are therefore totally unaware of this step. In the case of crash
failures, only the first copy of each message that is received by
a~\ac{SE} is processed; all further copies are dropped by the
receiver. In the case of Byzantine failures with replication level
$M=2f+1$, each entity must wait for at least~$(f+1)$ copies of the same
message before it can handle it. Once a strict majority has been
reached, the message can be processed and all further copies of the
same messages that might arrive later on can be dropped.

\paragraph*{Entities Migration}
\ac{PADS} can benefit from the migration of~\acp{SE} to balance
computation/communication load and reduce the communication cost, by
placing the~\acp{SE} that interact frequently ``next'' to each other
(e.g. on the same~\ac{LP})~\cite{gda-simpat-2017}. In FT-GAIA, the
entity migration is subject to a new constraint: the instances of the
same~\ac{SE} can never reside on the same~\ac{LP}. More specifically,
the~\acp{SE} migration is handled by the underlying GAIA/ART\`IS
middleware: each~\ac{LP} runs a clustering mechanism based on a
heuristic function
that tries to put together (on the same~\ac{LP}) the~\acp{SE} that
interact frequently through message exchanges. Special care is taken
to avoid putting too many entities on the same~\acp{LP} that would
become a bottleneck. Once a new feasible allocation is found, the
migration of a~\ac{SE} is implemented through moving its state
variables to the destination~\ac{LP}.  In different terms, our design
choice has been to maintain GAIA and FT-GAIA as separate as
possible. In fact, the clustering heuristics used by GAIA are totally
unaware of the functional replication of~\acp{SE}. This has simplified
the development of FT-GAIA as a separate software module at the cost
of using the generic self-clustering heuristics provided by GAIA.
Most likely, specifically tailored heuristics would be able to obtain
a better clustering of~\acp{SE} when considering the presence of
copies of the same~\acp{SE}.

%% file: sec_experimental-evaluation.tex
\section{Experimental Performance Evaluation}\label{sec:experimental-evaluation}

In this section we evaluate a prototype implementation of FT-GAIA by
implementing a simple simulation model of a Peer-to-Peer (P2P) communication
system. The simulation model built on top of~FT-GAIA is executed under
different workload parameters that will be described in the following.
The~\ac{WCT} of the simulation runs is recorded (excluding the time 
to setup the simulation) such as other metrics of interest.
The tests were performed on a cluster of workstations, each host
being equipped with an Intel Core i5-4590 3.30 GHz processor with
4 physical cores and 8~GB of RAM. The operating system was Debian Jessie. The workstations
are connected through a Fast Ethernet LAN.

\subsection{Simulation Model}

We simulate a simple~P2P communication protocol over randomly
generated directed overlay graphs. Nodes of the graphs are peers
while links represent communication connections~\cite{D'Angelo:2009,gda-jpdc-2017}.
In these overlays, all nodes have the same out-degree, that has been 
set to~$5$ in our experiments.  During the simulation, each node 
periodically updates its neighbor set.  Latencies for message transmission 
over overlay links are generated using a lognormal distribution~\cite{Farber:2002}.

The simulated communication protocol works as follows. Periodically,
nodes send~PING messages to other nodes, that in turn reply with a~PONG message that is used by the sender to estimate the average
latencies of the links (note that communication links are, in fact,
bidirectional). The destination of a PING is randomly selected to be a
neighbor (with probability $p$), or a non-neighbor (with probability
$1-p$). A neighbor is a node that can be reached through an outgoing
link in the directed overlay graph.

Each node of the P2P overlay is represented by a~\ac{SE} within
some~\ac{LP}. Unless stated otherwise, each~\ac{LP} was executed on a
different~\ac{PE}, so that no two~\acp{LP} shared the same CPU core.
Three different scenarios are considered: a \emph{no fault}
scenario, where no faults occur, a \emph{crash} scenario, where
crash failures occurs and finally a \emph{Byzantine} scenario
where Byzantine faults occurs.

We executed~$15$ independent replications of each simulation run. In most of
the charts in this section, mean values are reported with a~$99.5\%$ confidence
interval.

\subsection{Impact of the number of~\acp{LP} and \acp{SE}}

\begin{figure}[t]
  \centering\includegraphics[width=1.0\columnwidth]{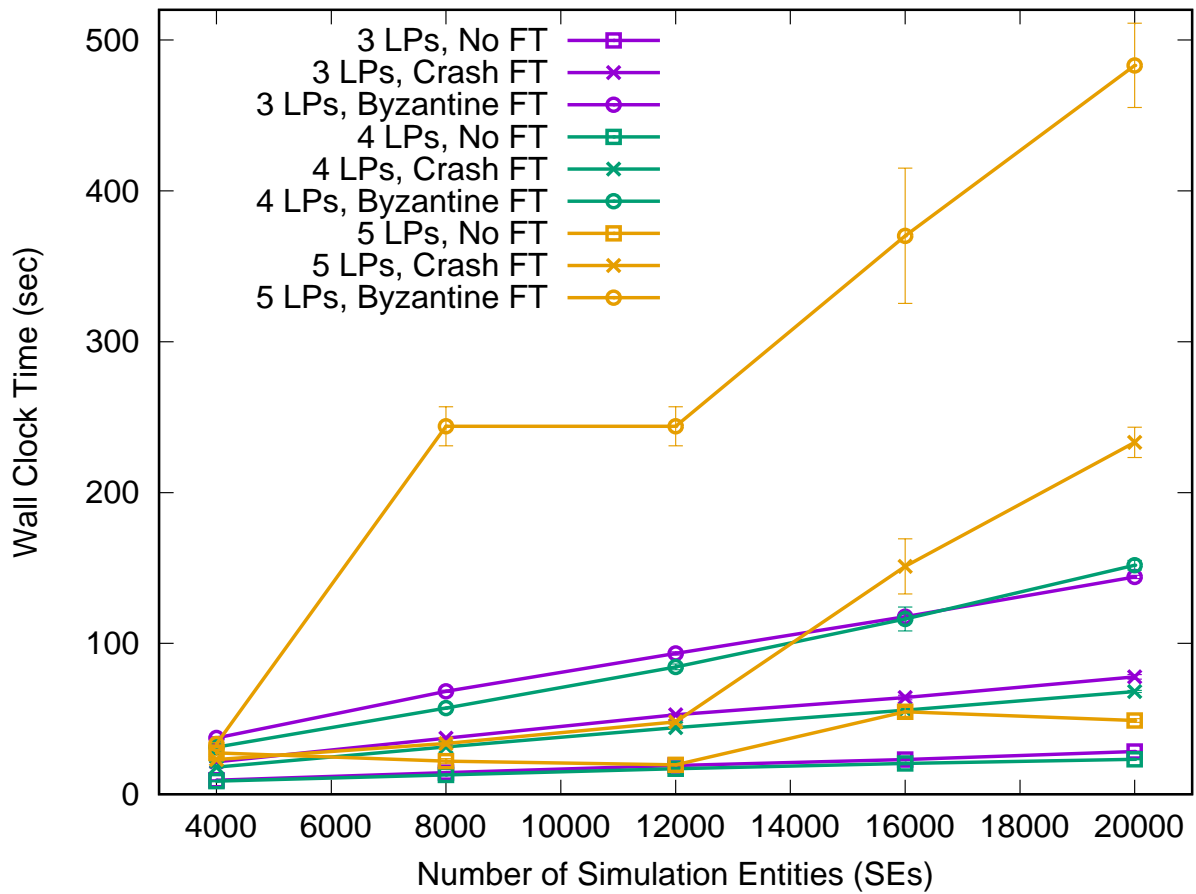}
  \caption{\acl{WCT} as a function of the number of~\acp{LP}, for
    varying number of~\acp{SE}. The number of~hosts is equal to the number of~\acp{LP}; migration is disabled. Lower is better.}\label{fig:diffENT}
\end{figure}

Figure~\ref{fig:diffENT} shows the~\ac{WCT} of the simulation that was
executed for~$10000$ timesteps with a varying number of~\acp{SE}; recall
that the number of~\acp{SE} is equal to the number of nodes in the~P2P
overlay graph. The number of~\acp{LP} was set to 3, 4, and 5; the number of hosts is equal to the number of~\acp{LP}, so that each~\ac{LP} is executed on a different physical machine. The~\ac{WCT} for the three failure scenarios is shown (i.e.,~no failure, 
single crash and single Byzantine failure). In all
cases, the adaptive migration heuristic provided by GAIA is disabled.

Results with~3 and~4 \acp{LP} are similar, with a slight improvement
with 4~\acp{LP}. Conversely, higher~\ac{WCT} is observed when
5~\acp{LP} are used.  As expected, the higher the number of~\acp{SE}
the higher the~\ac{WCT}. This happens since the simulation incurs in a higher
communication overhead. All curves show a similar trend: in
particular, it is worth noting that the increment due to the faults 
management schemes is mainly caused by the higher number of messages that are
exchanged among nodes.

\begin{figure}[t]
  \centering%
  \includegraphics[width=0.8\columnwidth]{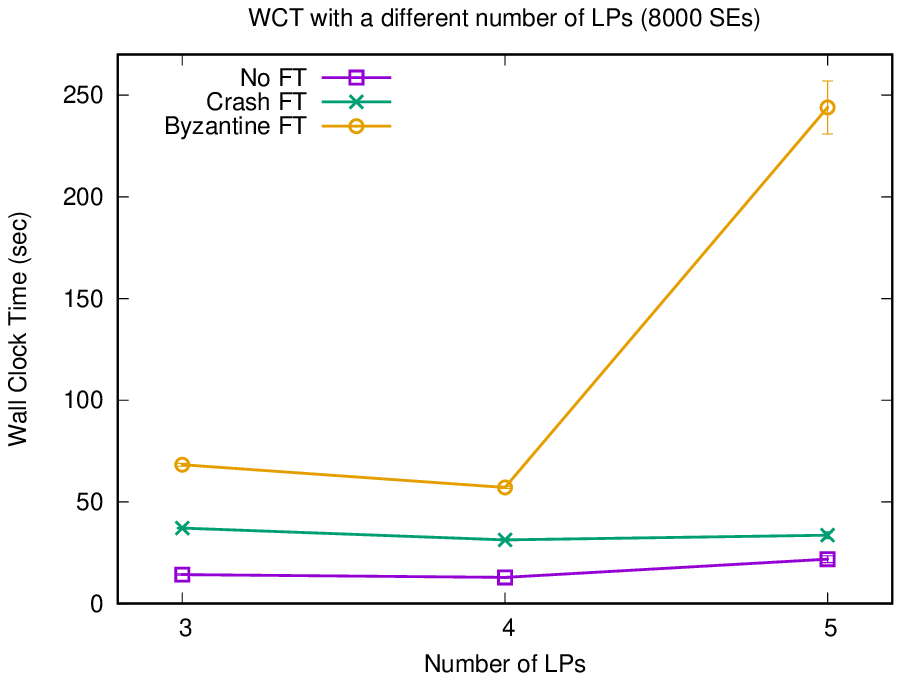}
  \caption{\acl{WCT} as a function of the number of~\acp{LP}, with
    8000~\acp{SE}; migration is disabled. Lower is better.}\label{fig:diffLP345_8k}
\end{figure}

\begin{figure}[t]
  \centering%
  \includegraphics[width=0.8\columnwidth]{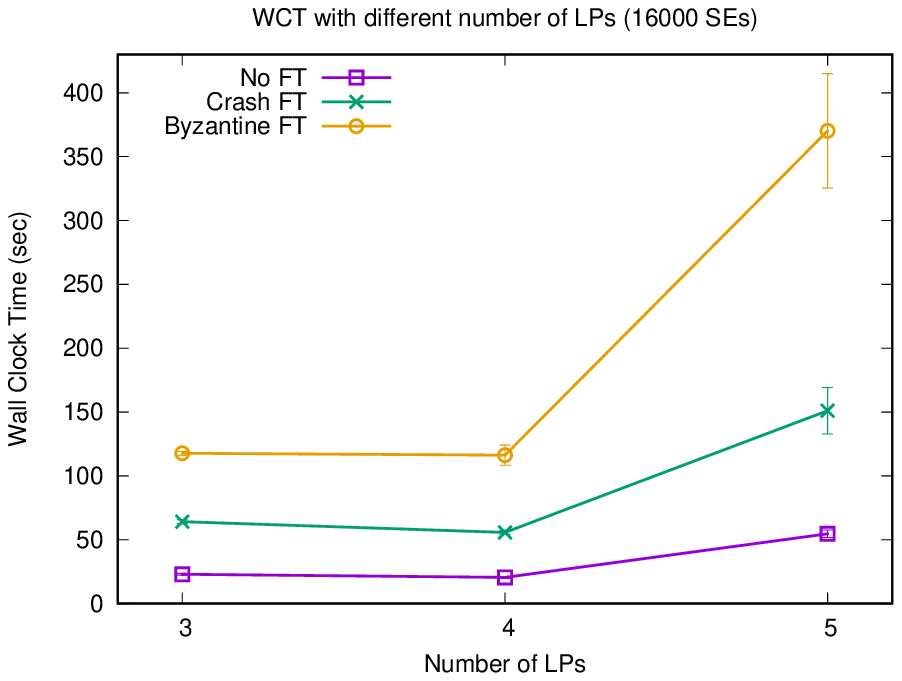}
  \caption{\ac{WCT} as a function of the number of~\acp{LP}, with
    16000~\acp{SE}. Migration is disabled. Lower is better.}\label{fig:diffLP345_16k}
\end{figure}

Figures~\ref{fig:diffLP345_8k} and~\ref{fig:diffLP345_16k} show
the~\ac{WCT} with~$8000$ and~$16000$ \acp{SE} with varying number of~\acp{LP};
again, each~\ac{LP} has been executed on a different physical host.
The two charts emphasize the increment of the time required to
complete the simulations with $5$~\acp{LP} and in presence of Byzantine
faults. This is due to the increased number of messages exchanged
among the~\acp{LP}: each message needs to be sent to three ($2M+1$) different
destinations in order to guarantee the expected fault tolerance.

\subsection{Impact of the number of LPs per host}

\begin{figure}[t]
  \centering%
  \includegraphics[width=1.0\columnwidth]{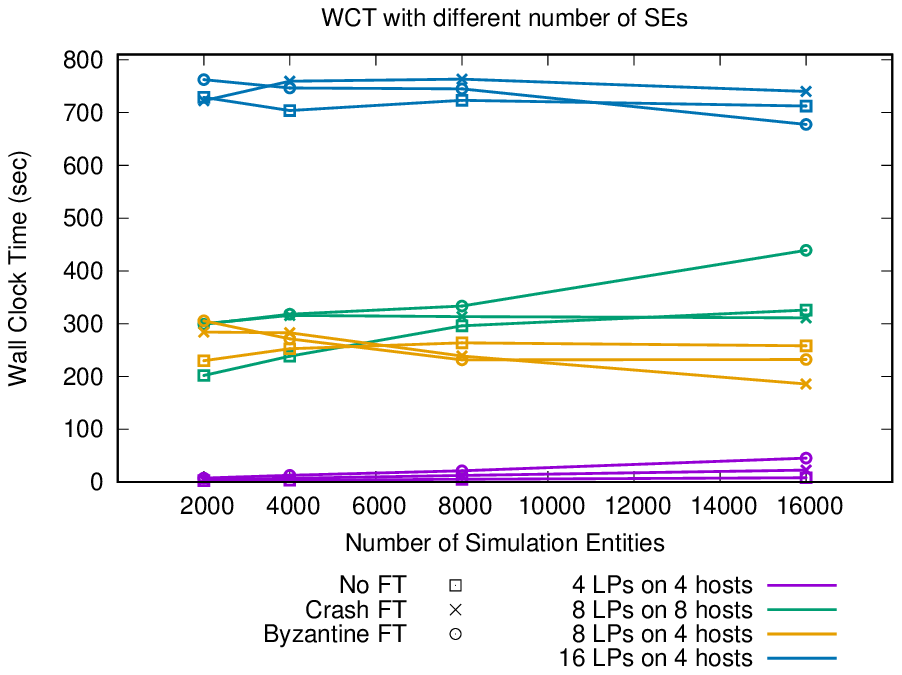}
  \caption{\ac{WCT} as a function of the number of~\acp{LP}, with
    different numbers of~\acp{LP} for each~host; migration is disabled. Lower is
    better.}\label{fig:diffENT_488b16q}
\end{figure}

In the previous experiments, each~\ac{LP} has been allocated in a
different host. Figure~\ref{fig:diffENT_488b16q} shows the~\ac{WCT}
when more than one~\ac{LP} is run in each host. In particular, the
following setups are considered: (\emph{i}) 4~\acp{LP} placed over
4~hosts (1~\ac{LP} per host), (\emph{ii}) 8~\acp{LP} placed over
8~hosts (1~\ac{LP} per host), (\emph{iii}) 8~\acp{LP} placed over
4~hosts (2~\acp{LP} per host), and (\emph{iv}) 16~\acp{LP} over
4~\acp{PE} (4~\acp{LP} per host). Note that, in any case, the number of~\acp{LP}/host never exceeds the number of cores/host, so that every~\ac{LP} runs on a separate processor core. For each setup, the
three failure scenarios already mentioned (no failures, crash,
Byzantine failures) are considered. Again, the migration heuristic provided by GAIA
is disabled. Each curve in the figure is related to one of those 
scenarios, when varying the amount of~\acp{SE}. It is worth noting that, 
when two or more~\acp{LP} are run on the same host, they can communicate
using shared memory rather than through the LAN. This means that, in this case the
inter-LP communication is more efficient.
For better readability, in this experiment the confidence intervals have 
been calculated but not reported in the figure.

We observe that the scenario with 4~\acp{LP} over 4~hosts is
influenced by the number of~\acp{SE} and the failure scenario, while in
the other cases it is the number of~\acp{LP} that mainly determines
the simulator performance.
When $8$~\acp{LP} are executed on $4$~hosts,
the performance is slightly better than the case where $8$~\acp{LP}
are executed on $8$~hosts. This is due to the better communication efficiency provided by shared memory with respect to the LAN interface.

The worst performance is measured when $16$~\acp{LP} are executed on
$4$~hosts. This is due to the fact that the amount of computation
in the simulation model is quite limited. Therefore, partitioning
the~\acp{SE} in $16$~\acp{LP} has the effect to increase the
communication cost without any benefit from the computational
point of view (i.e.,~in the model there is not enough computation
to be parallelized).

\subsection{Impact of the number of failures}

\begin{figure}[t]
  \centering%
  \includegraphics[width=.8\columnwidth]{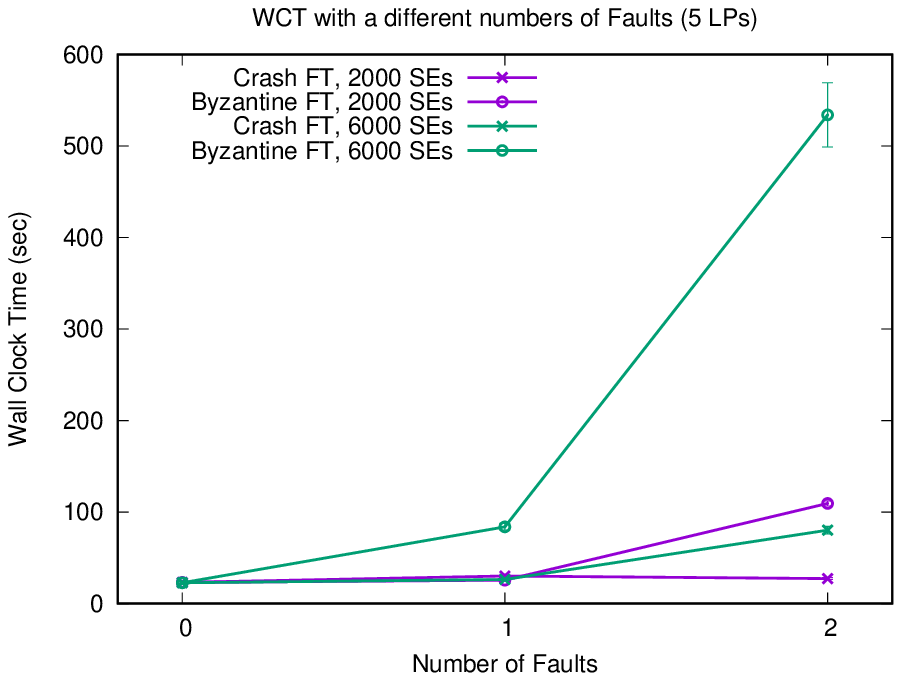}
  \caption{\ac{WCT} as a function of the number of faults; $10000$
    timesteps with $5$~\acp{LP}; migration is disabled. Lower is
    better.}\label{fig:diffMAXFT012}
\end{figure}

\begin{figure}[t]
  \centering%
  \includegraphics[width=.8\columnwidth]{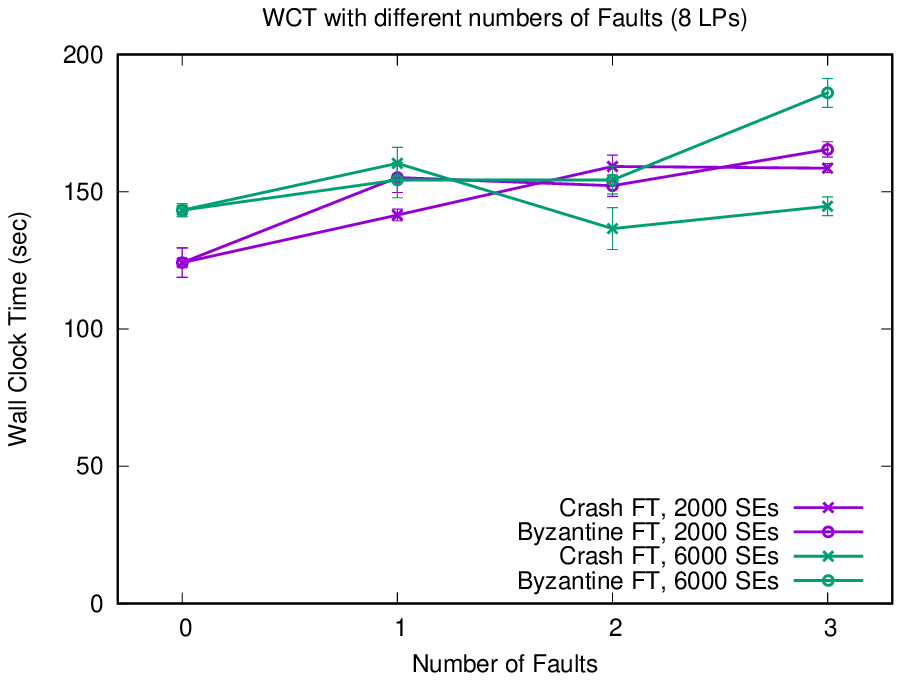}
  \caption{\ac{WCT} as a function of the number of faults; $2000$
    timesteps over $8$~LPs; migration is disabled. Lower is better.}\label{fig:diffMAXFT0123}
\end{figure}

The impact of the number of faults on the simulation \ac{WCT} is now studied. 
Two different setups are considered, one with $5$~\acp{LP} over $5$~hosts
(Figure~\ref{fig:diffMAXFT012}), and one with $8$~\acp{LP} over $4$~hosts
(Figure \ref{fig:diffMAXFT0123}).
The choice of $5$~\acp{LP} is motivated by the fact that this is the minimum
number of~\acp{LP} that allows us to tolerate up to two Byzantine faults. 
Furthermore, the~P2P simulation model used in this performance evaluation shows
a significant degradation of performance when the number of~\acp{LP} is larger than~$8$.
As described before, this is due to the specific characteristics of the simulation model, 
in which there is a limited amount of computation that can be parallelized. 
On the other hand, partitioning the model on a large number of~\acp{LP} sharply 
increases the communication cost.
More in detail, the setup with $8$~\acp{LP} on $4$ hosts allows testing $3$~Byzantine
faults with $2$~\acp{LP} per host in a setup with a limited communication overhead.

Figure~\ref{fig:diffMAXFT012} shows the WCTs measured with $0$, $1$ and~$2$
faults. Each curve refers to a scenario with~$2000$ or~$6000$ \acp{SE}
with crash or Byzantine failures. As expected, the higher the number
of faults, the higher the~\acp{WCT}, especially when Byzantine faults
are considered. Indeed, in this case a higher amount of communication
messages is required among~\acp{SE} in order to properly handle 
faults.

A higher~\ac{WCT} is measured with 8~\acp{LP}, as shown in
Figure~\ref{fig:diffMAXFT0123}. In this case, the amount of faults
has a limited influence on the simulation performance.
As before, the computational load of this simulation model is too low
for gaining from the partitioning in $8$~\acp{LP}. 
In other words, the latency introduced by network communications is so high
that both the number of~\acp{SE} and the number of faults have a negligible impact on
performances.

\subsection{Impact of~\acp{SE} migration}

\begin{figure}[t]
  \centering%
  \includegraphics[width=.8\columnwidth]{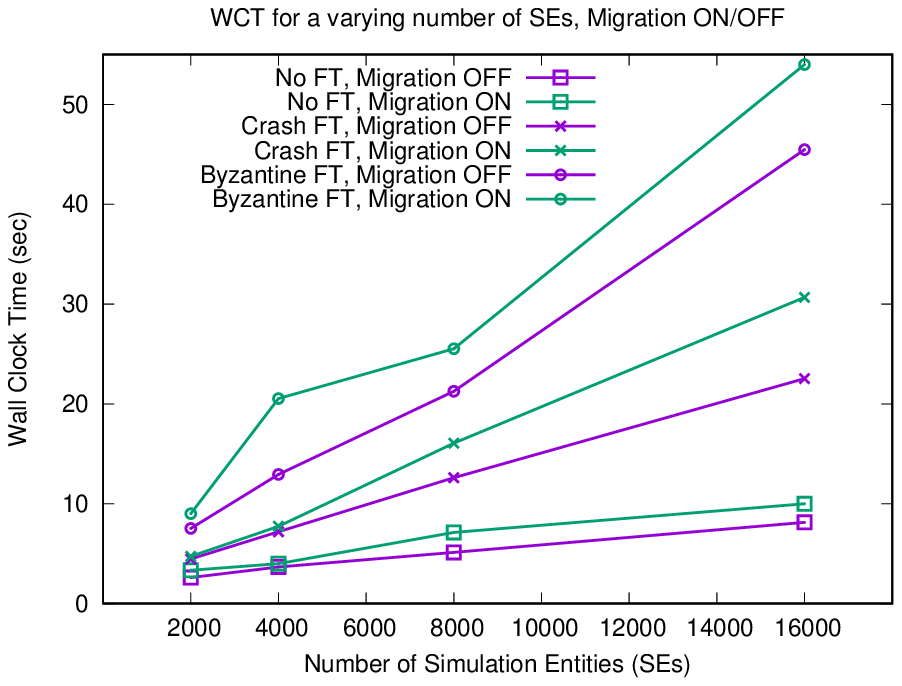}
  \caption{\ac{WCT} with~\acp{SE} migration ON/OFF, as a function of the
    number of~\acp{SE}. Lower is better.}\label{fig:diffMIGR}
\end{figure}

Finally, Figure~\ref{fig:diffMIGR} shows the~\ac{WCT} of a simulation composed of 4~\acp{LP} 
(in which each~\ac{LP} was executed on a different host) with different failure
schemes, when the adaptive migration of~\acp{SE} provided by the GAIA framework 
is enabled/disabled. Also in this case, for better readability, the confidence 
intervals are not reported in figure.

In this case, the trend obtained with the~\acp{SE} migration is similar to that
obtained when no migration is performed but the overall performance
are better when the migration is turned off. This is due to the
overhead introduced by the self-clustering heuristics and the state of
the~\acp{SE} that are transfered between the~\acp{LP}. In other
words, the adaptive clustering of~\acp{SE} that in many other simulation models
has provided a significant gain, in this case, is unable to give a speedup.

The main motivation behind this result is the fact that, in this prototype, 
we have decided to use the very general clustering heuristics that are already 
implemented in GAIA/ART\`IS. These heuristics assume that the simulation model
is composed of a set of agents, each one with its specific behavior and
communication pattern. In the case of FT-GAIA, this not true. In fact, all the
copies of a given \ac{SE} share exactly the same behavior and interactions.
Moreover, as described before, FT-GAIA adds the constraint that the instances of
the same~\ac{SE} can never reside on the same~\ac{LP}. This constraint affects the free
flow on~\acp{SE} among the~\acp{LP} and consequently reduces the clustering
efficiency.

For these reasons, we think that more specific replication-aware clustering
heuristics need to be designed to improve the clustering performance
while balancing the overhead introduced by the fault tolerance mechanism.

%% file: sec_analytical-evaluation.tex
\section{Analytical Reliability Evaluation}\label{sec:analytical-evaluation}

In Section~\ref{sec:fault-tolerant-simulation} we have seen that
the~FT-GAIA extension of the GAIA/ART\`IS middleware works by
making~$M$ copies of each~\ac{SE}, and ensuring that each
copy resides on a different~\ac{LP}. This requirement, that we call
\emph{FT-GAIA constraint} from now on, guarantees that FT-GAIA can
tolerate up to~$M-1$ crash failures of~\acp{LP}, or up to $\lfloor
(M-1)/2 \rfloor$ Byzantine failures.

In this section we perform a reliability analysis of an~FT-GAIA
simulation to complement the experimental performance evaluation from
Section~\ref{sec:experimental-evaluation}. The goal of this analysis
is to estimate the reliability of~FT-GAIA when the number of failures
is higher than the thresholds above; also, we want to study what
happens if the~FT-GAIA constraint is not enforced, that is, what
happens if more than one instance of the same simulation entity is
allowed reside on the same~\ac{LP}. These kinds of analyses would be
complex and time-consuming if performed through actual experiments as
in the previous section, so we resort to a simpler probabilistic
evaluation. We remark that the analysis below is only
  concerned with the system reliability, and does not consider any
  performance metric. Indeed, the content of this section is orthogonal
  to the performance analysis described in
  Section~\ref{sec:experimental-evaluation}. Analytical performance
  models for distributed simulations have been proposed in the
  past~\cite{Quaglia:1999}, but their extension to FT-GAIA would be
  non-trivial and is outside the scope of this work.

We analyze the system reliability of FT-GAIA under crash or Byzantine
failures of the~\acp{LP}, since they are the basic component that can
fail in GAIA-FT. Indeed, a crash of a whole host implies a crash of
all the~\acp{LP} running on it, and a crash of a~\ac{SE} implies a
crash of the whole~\ac{LP} where the~\ac{SE} is executed.

The analysis presented below relies on the following assumptions:

\begin{itemize}
  \item All crashes are permanent: a crashed~\ac{LP} is never brought
    back to a functioning state.
  \item Every~\ac{LP} has the same probability to crash. 
  \item All instances of each simulation entity are randomly and
    uniformly placed on the available~\acp{LP}, either respecting or
    not respecting the FT-GAIA constraint (we will analyze both
    scenarios).
  \item \acp{SE} are never migrated from one~\ac{LP} to
    another.
\end{itemize}

While some of the assumptions above are quite limiting, they simplify
the analysis considerably and still provide useful qualitative
information.

\subsection{Crash Failure Model}

Given a simulation with~$L$ \acp{LP} and~$N$ simulation entities,
with~$M$ instances of each entity ($1 \leq M \leq L$), we assume
that~$X$ randomly chosen \acp{LP} crash during the simulation ($0 \leq
X \leq L$). We want to compute the system reliability, that is, the
probability that a sufficient number of instances of each entity
survived to ensure that the simulation produces the intended
results. In the crash failure model, the reliability~$R_C$ is the
probability that at least one instance of each entity resides on
a~\ac{LP} that does not crash; in case of byzantine failures, the
reliability~$R_B$ is the probability that at least $\lceil (M+1)/2
\rceil$ entities (the majority) reside on~\acp{LP} that do not crash.

For each~\ac{SE}~$i$, let~$N_i$ be the random variable denoting the
number of instances of~$i$ that reside on \acp{LP} that did not crash.
The pmf (probability mass function) $\Pr(N_i = k)$, $0 \leq k \leq M$,
can be derived easily by casting the original problem into an ``urn
problem''. If~$k$ is greater than~$L-X$, then $\Pr(N_i = k)$ is zero
since less than~$k$ \acp{LP} survived through the end of the
simulation. If $0 \leq k \leq L-X$, then $P(N_i = k)$ is the
probability of getting~$k$ white balls out of~$M$ extracted without
replacement from an urn containing~$X$ black balls (representing
crashed~\acp{LP}) and~$L-X$ white balls (representing~\acp{LP} that
did not crash). Therefore we have:

\begin{align}
\Pr(N_i = k) &= \begin{cases}
  \displaystyle \binom{X}{M-k} \binom{L-X}{k} / \binom{L}{M} & \mbox{if $0 \leq k \leq L-X$}\\
  0 & \mbox{if $L-X < k \leq M$}
  \end{cases}
\end{align}

The system reliability~$R_C$ under the crash failure model is the
probability that the simulation terminates successfully. This is the
joint probability that $N_i \geq 1$ for each~$i$. If there are more
instances of each~\ac{SE} than crashed~\acp{LP}, then~$R_C=1$ since
the FT-GAIA constraint ensures that there is at least one live
instance of each entity. On the other hand, if $M \leq X \leq L$ it
may happen that all instances of the same entity fail, and the system
reliability can then be computed in this case as:


\begin{align*}
\prod_{i=1}^N \Pr(N_i \geq 1) &= \prod_{i=1}^N \left(1 - \Pr(N_i = 0) \right) = \left[ 1 - \displaystyle \binom{X}{M} / \binom{L}{M} \right]^N
\end{align*}

\noindent Therefore, $R_C$ is defined as:

\begin{align}
  R_C &= \begin{cases}
    1 & \mbox{if $0 \leq X < M$}\\
    \left[ 1 - \displaystyle \binom{X}{M} / \binom{L}{M} \right]^N & \mbox{if $M \leq X \leq L$} 
  \end{cases} \label{eq:reliability:gaiaft}    
\end{align}

Note that if~$X = L$ (all~\acp{LP} failed) then~$R_C$ is zero as
expected. Also, observe that~$R_C$ tends to zero as the number of
entities~$N$ approaches infinity.

\subsection{Byzantine Failure Model}

The reliability~$R_B$ under the Byzantine failure model can be
computed in a similar way. The minimum number of working instances of
each~\ac{SE} that are required to guarantee that the simulation
terminates is $\lceil (M+1)/2 \rceil$. If the number of failures~$X$
is strictly lower than $\lceil (M+1)/2 \rceil$, then~$R_B = 1$. If the
number of failed~\acp{LP} is greater than or equal to $\lceil (M+1)/2
\rceil$, the reliability becomes strictly less than~$1$ and can be
computed as the joint probability that the majority of the instances
of each entity~$i$ are active:

\begin{align*}
\prod_{i=1}^N \Pr(N_i \geq \lceil (M+1)/2 \rceil) &= \prod_{i=1}^N \left[ \sum_{k=\lceil (M+1)/2 \rceil}^L \Pr(N_i = k) \right] \nonumber\\
    &= \left[ \displaystyle \sum_{k=\lceil (M+1)/2 \rceil}^L \Pr(N_i = k) \right]^N
\end{align*}

\noindent Hence we have:

\begin{align}
  R_B &= \begin{cases}
    1 & \mbox{if $0 \leq X < \lceil (M+1)/2 \rceil$}\\
    \left[ \displaystyle \sum_{k=\lceil (M+1)/2 \rceil}^L \Pr(N_i = k) \right]^N & \mbox{if $\lceil (M+1)/2 \rceil \leq X \leq L$} 
  \end{cases}\label{eq:reliability:byzantine}
\end{align}

\begin{figure}
  \centering\includegraphics[width=1.0\textwidth]{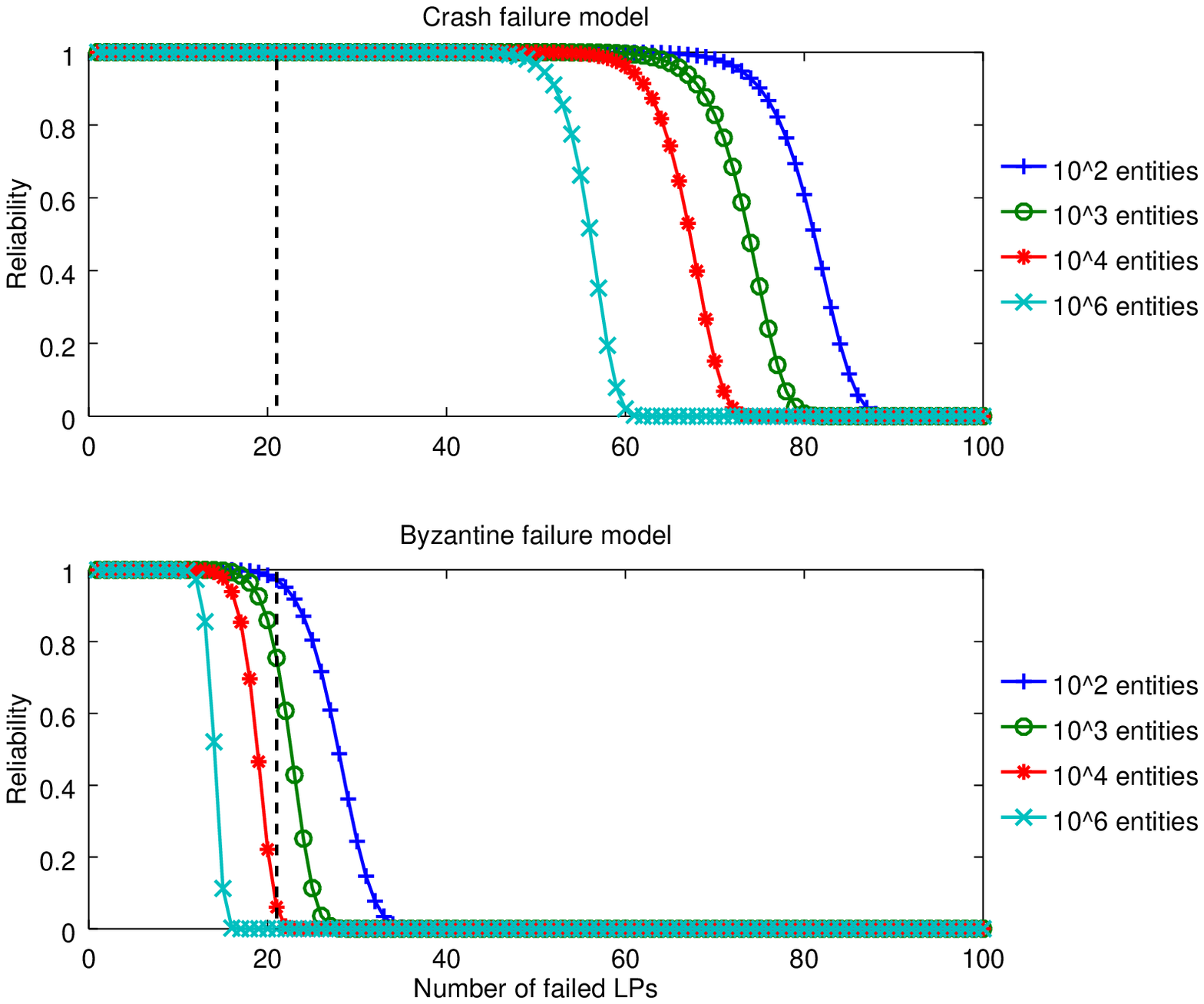}
  \caption{Reliability of FT-GAIA to crash (top) and Byzantine
    failures (bottom), as a function of the number of failures~$X$; we
    assume~$L=100$~\acp{LP} and~$M=21$ instances of each entity. The
    vertical line is at~$M$.}\label{fig:reliability-X}
\end{figure}

Figure~\ref{fig:reliability-X} shows the reliability of~FT-GAIA using
$L=100$~\acp{LP} with~$M=21$ instances of each entity, as a function
of the number of crashes~$X$. Under the crash failure model (top
figure) the system tolerates up to~$M-1 = 20$ crashes; under the
Byzantine failure model (bottom figure), the system tolerates up to
$\lceil (M+1)/2 \rceil - 1 = 10$ crashes. When~$X$ exceeds the
thresholds, the reliability drops; in fact, $R_B$ drops faster
than~$R_C$, because the Byzantine failure model requires a higher
number of active instances to guarantee that the simulation terminates
successfully.

\begin{figure}
  \centering\includegraphics[width=1.0\textwidth]{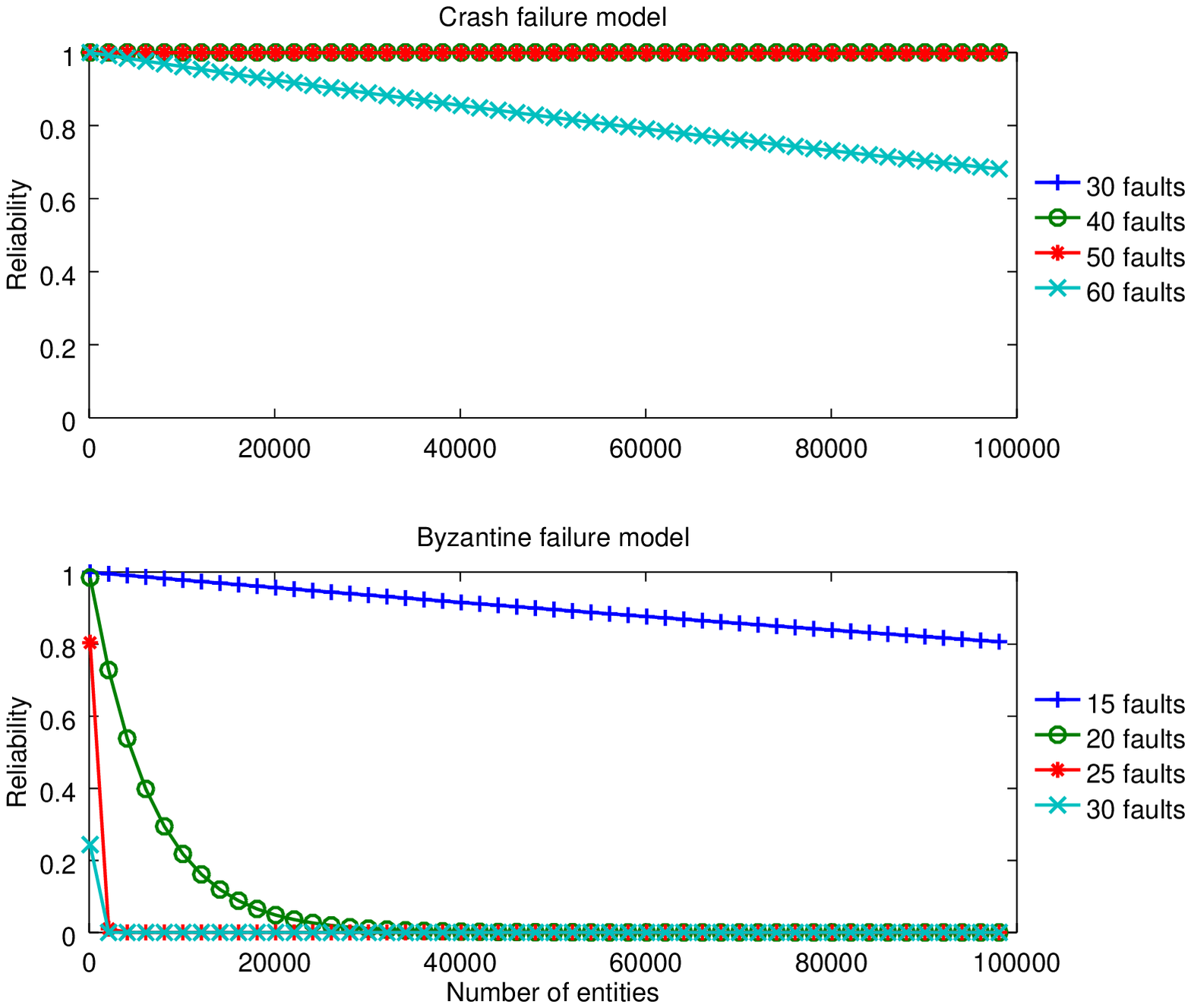}
  \caption{Reliability of FT-GAIA for crash failures as a function of
    the number of simulation entities~$N$, with~$L=30$ LPs and~$M=11$
    instances of each entity.}\label{fig:reliability-N}
\end{figure}

Figure~\ref{fig:reliability-N} shows the reliability of~FT-GAIA as a
function of the number of entities~$N$ for different number of
faults~$X$ (note that the values of $X$ differ for the crash and
Byzantine failure models); we assume $L=100$ \acp{LP} and $M=21$
instances of each entity. Protecting the simulation against Byzantine
faults requires a higher number of active instances for each~\ac{SE},
since the model is more general than the crash failure model. However,
the drawback is that the reliability $R_B$ drops very quickly as~$N$
increases even when the number of faults~$X$ slightly exceeds the
threshold. Therefore, the user must be aware that Byzantine faults are
much more sensitive to the choice of the ``correct'' value of~$M$ than
crash failures.

\subsection{Impact of the FT-GAIA Constraint}

We now study what would happen if the~FT-GAIA constraint is not
applies, i.e., if~FT-GAIA were allowed to put more than one instance
of same entity on the same~\ac{LP}.  Given a simulation with
$L$~\acp{LP}, $N$ entities that are replicated~$M$ times, and
$X$~\acp{LP} that crash during the simulation, let~$N^*_i$ be the
number of surviving instances of entity~$i$ under the assumption that
the~FT-GAIA constraint does not apply. This scenario can again be
analyzed as an urn problem, in this case where the balls are extracted
with replacement. The random variables~$N^*_i$ follow a binomial
distribution $B\left(M, \frac{L-X}{L}\right)$, so we have:

\begin{align*}
\Pr(N^*_i = k) &= \displaystyle \binom{M}{k} \left( \frac{L-X}{L} \right)^k \left(\frac{X}{L} \right)^{M-k}
\end{align*}

As above, the system reliability~$R^*_C$ under the crash failures
model can be expressed as:

\begin{align}
  R^*_C &= \prod_{i=1}^N \Pr(N^*_i \geq 1)
  = \prod_{i=1}^N \left( 1 - \Pr(N^*_i = 0) \right) 
  = \left[ 1 - \left( \frac{X}{L} \right)^M \right]^N \label{eq:reliability:nogaiaft}
\end{align}

Eq.~\eqref{eq:reliability:nogaiaft} tells us that the system
reliability~$R^*_C$ is strictly less than~$1$ even in presence of a
single crash failure. Indeed, if the instances of each~\ac{SE} are
randomly placed on the~\acp{LP}, there is a small but non-negligible
probability that all instances of, say, entity~$i$ are placed on the
same~\ac{LP} that will crash, aborting the whole simulation. This can
not happen if the~FT-GAIA constraint is enforced.

\begin{figure}
  \centering\includegraphics[width=\textwidth]{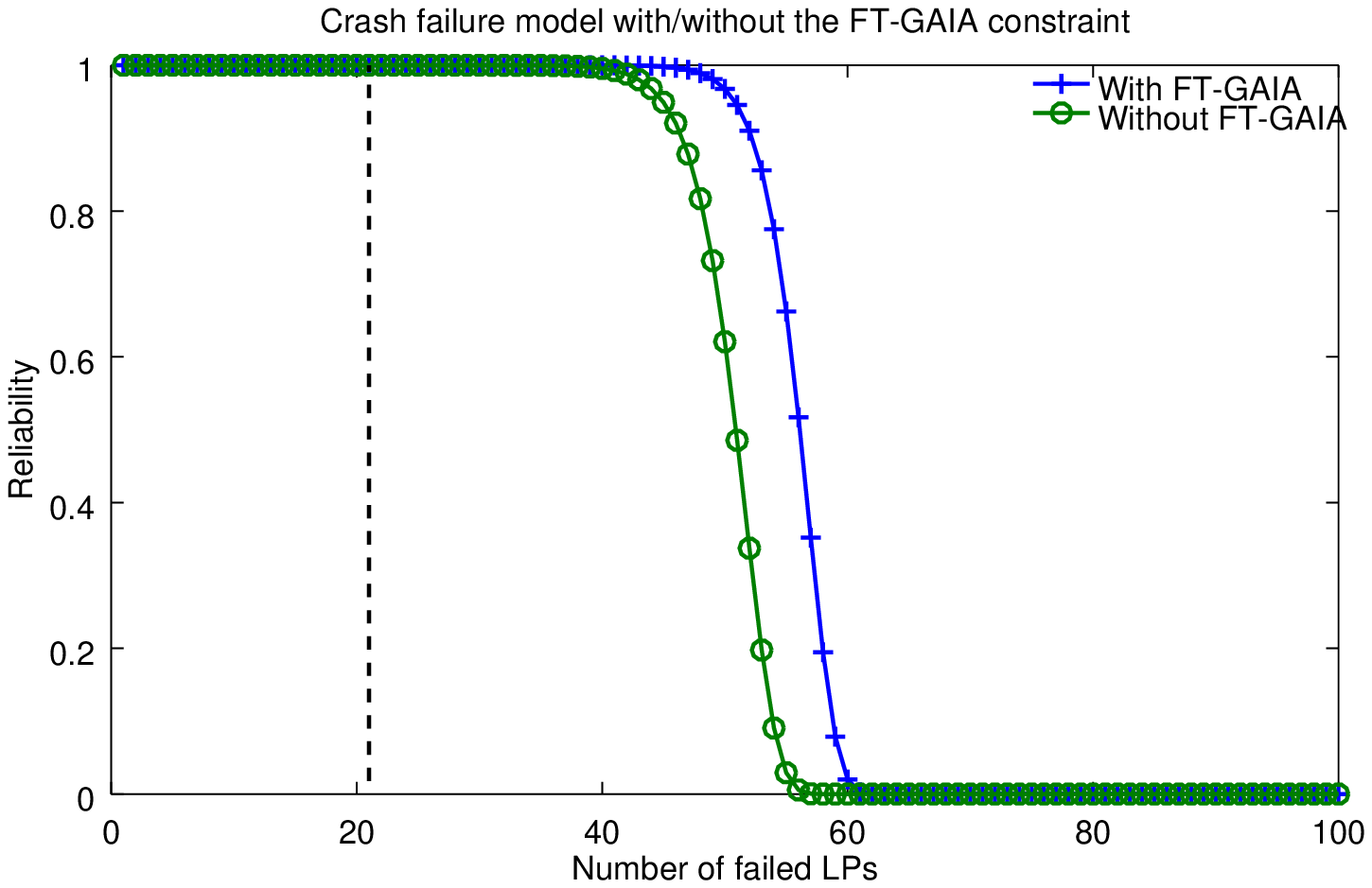}
  \caption{Reliability with and without the FT-GAIA constraint as a
    function of the number of failed LPs~$X$, with~$L=100$ LPs,
    $N=10^6$ entities and~$M=21$ instances of each entity (vertical
    line).}\label{fig:reliability-ft-noft}
\end{figure}

Figure~\ref{fig:reliability-ft-noft} compares the system reliability
with and without the~FT-GAIA constraint. We consider a system with
$L=100$~\acp{LP} and~$N=10^6$ simulation entities that are
replicated~$M=21$ times. The FT-GAIA constraint allows the system to
sustain up to $M-1 = 20$ failures; indeed, when $X < M$ the
reliability~$R_C$ computed using Eq.~\eqref{eq:reliability:gaiaft}
is~$1$. When $X < M$ the reliability $R^*_C$ computed using
Eq.~\eqref{eq:reliability:nogaiaft} is slightly less than~$1$;
however, the difference is so tiny to be almost negligible. Indeed,
Eq.~\ref{eq:reliability:nogaiaft} shows that the probability that all
instances of one~\ac{SE} reside on the same (crashed) \ac{LP} gets
smaller as the number of replicas~$M$ increases. However, it is
important to remember that this is true if the~\ac{SE} instances are
randomly placed on the~\acp{LP}.

In practice, however, the placement is \emph{not} random, at least
when the automatic clustering and migration facilities of~GAIA/ART\`IS
are enabled. Indeed, GAIA/ART\`IS monitors the communication pattern
of the~\acp{SE}, and migrate those that exhibit a high level of
interaction on the same~\ac{LP} to reduce the number of remote
communications~\cite{gda-simpat-2017}. If the placement of~\acp{SE} is
not random, the FT-GAIA constraint becomes essential to limit the
probability that too many instances of the same~\ac{SE} fail at the
same time.


  \subsection{Discussion}

  We can use the results above to provide some guidelines on how the
  replication level~$M$ can be chosen in practice. Note that choosing
  the ``best'' value of~$M$ is a difficult problem, since the answer
  depends on the simulation model that is executed, on the execution
  environment, and on the failure model that is considered.

  If the user requests a strong guarantee that the simulation run is
  completed without failures, then it is necessary to choose a value
  of~$M$ that produces a system reliability equal to~$1$.  Assuming
  that the GAIA-FT constraint is enforced,
  Eq.~\ref{eq:reliability:gaiaft} and~\eqref{eq:reliability:byzantine}
  tells us that the system reliability is one if the number of
  expected failures~$X$ is strictly less than~$M$ for the crash
  failure model, and strictly less than $\lfloor (M+1)/2 \rfloor$ for
  the Byzantine failure model.

  The number of expected failures~$X$ can be expressed as
  
  \begin{align}
    X &= L \lambda t \label{eq:X}
  \end{align}
  
  \noindent where $\lambda$ is the failure rate of each~\ac{LP},
  and~$t$ is the duration of the simulation run. Both parameters can
  be estimated empirically; in particular, $\lambda$ can be computed
  as the inverse of the~\ac{MTTF}, that is a quantity that can be
  easily observed from the operational history of the system.

  Therefore, the simulation can be completed with probability~$1$ in
  the crash failure model if~$X < M$; taking into account
  Eq.~\eqref{eq:X} we get:
  \begin{align}
    M > L \lambda t \label{eq:M:crash}
  \end{align}

  Similarly, the simulation can be completed with probability~$1$ in
  the Byzantine failure model if $X < \lfloor (M+1)/2 \rfloor$; again,
  taking into account Eq.~\eqref{eq:X} we get:
  \begin{align}
    M > 2 L \lambda t - 1 \label{eq:M:byzantine}
  \end{align}

  The user is responsible for deciding which failure model to
  use. Once the choice is made, the smallest integer value~$M$
  satisfying~\eqref{eq:M:crash} or~\eqref{eq:M:byzantine} is the
  replication level that provides the strongest guarantee to complete
  the simulation, under the simplifying assumptions stated at the
  beginning of this section.

  The experimental evaluation illustrated in
  Section~\ref{sec:experimental-evaluation} shows that providing
  protection against Byzantine failures is more costly in term of wall
  clock time; however, Byzantine failures are more general than crash
  failures. If the user trusts the computation and assumes that a
  running~\ac{SE} will always compute the correct result, the more lax
  crash failure model can be considered, allowing a lower replication
  level~$M$ to be chosen.

%% file: sec_conclusions.tex
\section{Conclusions and Future Work}\label{sec:conclusions}

In this paper we described an approach to provide fault tolerance through functional 
replication in parallel and distributed simulations.
Our solution, called FT-GAIA, is an extension to the
GAIA/ART\`IS simulation middleware that acts transparently to the user that 
creates and manages the simulation.
Fault tolerance is provided by replicating simulation entities and distributing 
them on multiple execution nodes. 
This is a particularly important issue to cope with, especially if we expect 
to have execution nodes running complex simulation over virtual machines hosted by public or private cloud systems.
Replication of their execution guarantees
tolerance to crash-failures and Byzantine faults of computing nodes.
In order to mitigate the costs of communication among simulation entities,
the middleware exploits an automatic migration of simulated entities among 
execution nodes with the aim to balance the computational load and minimize 
the communication overhead.

A preliminary performance evaluation of FT-GAIA has been presented, based on a
prototype implementation. Results show that a high degree of fault
tolerance can be achieved, at the cost of a moderate increase in the
computational load of the execution units. Moreover, a probabilistic model that
drives an analytical evaluation of the proposed scheme is introduced.

As a future work, we aim at improving the efficiency of FT-GAIA by leveraging 
on ad-hoc clustering heuristics that are aware of the fault tolerance
mechanism implemented by FT-GAIA. For example, evaluating the impact on the
clustering of all the copies of a given simulation entity instead of 
considering each entity by itself. Indeed, we believe that specifically
tuned clustering and load balancing mechanisms can significantly reduce 
the overhead introduced by the replication of the simulated entities.
Another aspect that needs to be investigated is the impact of the
functional replication on different synchronization algorithms used in
distributed simulations, e.g.~the Chandy-Misra-Bryant (CMB)
conservative approach based on NULL messages~\cite{cmb}, or the Time Warp
optimistic protocol~\cite{Jefferson85} based on rollbacks, that are
the most commonly used in practice.